\documentclass[letterpaper]{aa}
\usepackage{graphicx}
\usepackage{txfonts}

\begin{document}

\title{\object{IGR J17544$-$2619}: A new supergiant fast X-ray transient
revealed by optical/infrared observations\thanks{Based on observations
collected at the European Southern\- Ob\-ser\-va\-to\-ry, Chile, under proposal
ESO 71.D-0073.}}

\author{L.~J. Pellizza\inst{1} \and S. Chaty\inst{1} \and I.
Negueruela\inst{2}}

\institute{AIM - Astrophysique Interactions Multi-\'echelles (Unit\'e Mixte de
Recherche 7158 CEA/CNRS/Universit\'e Paris 7 Denis Diderot), CEA Saclay,
DSM/DAPNIA/Service d'Astrophysique, B\^at. 709, L'Orme des Merisiers, FR-91191
Gif-sur-Yvette Cedex, France \and Dpto. de F\'{\i}sica, Ingenier\'{\i}a de
Sistemas y Teor\'{\i}a de la Se\~nal, Universidad de Alicante, Apdo. 99, E03080
Alicante, Spain}

\offprints{L.J. Pellizza, \email{lpellizz@discovery.saclay.cea.fr}}

\date{Received / Accepted}

\abstract{One of the most recent discoveries of the {\em INTEGRAL} observatory
is the existence of a previously unknown population of X-ray sources in
the inner arms of the Galaxy. \object{IGR J17544$-$2619},
\object{IGR~J16465$-$4507} and \object{XTE~J1739$-$302} are among these
sources. Although the nature of these systems is still unexplained, the
investigations of the optical/NIR counterparts of the two last sources,
combined with high energy data, have provided evidence of them being highly
absorbed high mass X-ray binaries with blue supergiant secondaries and
displaying fast X-ray transient behaviour. In this work we present our
optical/NIR observations of \object{IGR J17544$-$2619}, aimed at identifying
and characterizing its counterpart. We show that the source is a high mass
X-ray binary at a distance of 2--4~kpc with a strongly absorbed O9Ib
secondary, and discuss the nature of the system.
\keywords{X-rays: binaries -- X-rays: individual: IGR J17544-2619}}

\maketitle

\section{Introduction}
\label{intro}

In recent years, a set of new X-ray sources were discovered by {\em INTEGRAL}
and {\em RXTE} observatories within a few tens of degrees of the direction to
the galactic center (Negueruela \cite{Neg04}; Kuulkers \cite{Kuu05}). These
sources present hard X-ray spectra, usually interpreted as due to strong
absoption by large hydrogen column densities arising from circumstellar
material. Their continuum spectral parameters are typical of neutron stars or
black holes. They are suspected to be high mass X-ray binaries (HMXBs) embedded
in highly absorbing media, and for some of them massive companions were
identified (e.g., Filliatre \& Chaty \cite{Fil04}; Smith \cite{Smi04};
Masetti et~al. \cite{Mas05}; Negueruela et~al. \cite{Neg05,Neg06}).
Among these sources there are transient ones, characterized by very short
outbursts ($\sim$hours) separated by large quiescence periods (Smith et~al.
\cite{Smi06}; in't~Zand \cite{Zan05}). \object{IGR~J16465$-$4507} and
\object{XTE~J1739$-$302}/\object{IGR~J17391$-$3021} are two of these transient
sources, which have been shown to be HMXBs with highly reddened early type
supergiant secondaries (Smith \cite{Smi04}; Negueruela et~al.
\cite{Neg05,Neg06}), hence supporting the hypothesis of these sources being
surrounded by large amounts of material.

\object{IGR J17544$-$2619} is a fast transient source lying in the direction of
the galactic center ($l = 3.24\degr$, $b = -0.34\degr$). It was discovered by
the IBIS/ISGRI instruments onboard {\em INTEGRAL} on September 17, 2003
(Sunyaev et~al. \cite{Sun03}), during an outburst which lasted around 2 hours.
A second outburst lasting for 8 hours was observed the same day (Grebenev
et~al. \cite{Gre03}), and a third one lasting for 10 hours was detected on March 8,
2004 (Grebenev et~al. \cite{Gre04}), demonstrating that this is a recurrent
transient source. An oscillation-like behaviour with a time scale of 1.5--2
hours has also been detected in the March 2004 observation (Grebenev et~al.
\cite{Gre04}). The source has also been observed by {\em XMM-Newton} on March
17 and September 11 and 17, 2003 (Gonz\'alez-Riestra et~al. \cite{GRi04}) and
by {\em Chandra} on July 3, 2004 (in't Zand \cite{Zan05}). {\em XMM-Newton}
observations revealed a hard spectrum, that can be modelled by an absorbed
power law with either variable absorption or variable spectral index, in both
cases with a large hydrogen column density ($N_{\mathrm{H}} \sim 1.9$--$4.3
\times 10^{22}$~cm$^{-2}$). {\em Chandra} observations also showed a hard
spectrum during bursts, but a softer one in quiescence, with similar absorption
($N_{\mathrm{H}} = 1.36 \pm 0.22 \times 10^{22}$~cm$^{-2}$). Assuming a
distance of 8~kpc, total unabsorbed luminosities of the source in the
0.5--10~keV range are of the order of $10^{32}$~erg~s$^{-1}$ in quiescence and
$10^{35}$--$10^{36}$~erg~s$^{-1}$ during activity. These values are typical of
HMXBs containing either neutron stars or black holes. {\em XMM-Newton} and {\em
Chandra} observations also allowed a precise positioning of the source
($4\arcsec$ and $0.6\arcsec$ error boxes respectively), challenging its
identification with the {\em ROSAT} source 1RXS J175428.3$-$2620 proposed by
Wijnands (\cite{Wij03}), and supporting the optical/NIR counterpart candidate
(\object{USNO-B1.0~0636-0620933}/\object{2MASS~J17542527$-$2619526}) suggested
by Rodriguez (\cite{Rod03}). The {\em XMM-Newton} ultraviolet magnitudes,
combined with optical/NIR data are also consistent with
\object{IGR~J17544$-$2619} having an early O-type companion (Gonz\'alez-Riestra
et~al. \cite{GRi04}).

One day after the discovery of \object{IGR J17544$-$2619}, we triggered our ESO
Target of Opportunity program to obtain optical and NIR images of the field of
this source, and optical spectra of the counterpart candidate proposed by
Rodriguez (\cite{Rod03}), which is also the brightest object in the {\em
XMM-Newton} error box. These measurements were aimed at confirming the
optical/NIR counterpart and determining its properties. In this paper we
present our observations (Sect.~\ref{obs}), their results (Sect.~\ref{res}),
and discuss the nature of the system (Sect.~\ref{disc}).

\section{Observations}
\label{obs}

Our observations of \object{IGR J17544$-$2619} were carried out on the nights
of September 18, 2003 (imaging) and May 11, 2004 (spectroscopy), with the ESO
3.5-meter New Technology Telescope (NTT) at La Silla Observatory, Chile.
Optical and NIR images of the field of the source were obtained with the ESO
Multi-Mode Instrument (EMMI) and the Son Of Isaac (SOFI) instruments
respectively, as part of a Target of Opportunity program (ESO 71.D-0073, P.~I.
Chaty). Intermediate resolution optical spectra of the counterpart candidate
proposed by Rodriguez (\cite{Rod03}) were also obtained with EMMI.

The red arm of EMMI, equipped with a mosaic of two $2048 \times 4096$ MIT CCD
detectors, was used in the longslit REMD mode together with grating \#6 and in
RILD mode with grisms \#3 and \#5 for spectroscopy. The same arm of EMMI was
used in RILD mode with $B$, $V$ and $R$ Bessel filters for imaging. The blue
arm, equipped with a Textronik TK1034 thinned, back-illuminated $1024 \times
1024$ CCD, was also used for spectroscopy in the longslit BLMD mode with
grating \#12. SOFI, equipped with a Rockwell Hg:Cd:Te $1024 \times 1024$ Hawaii
array, was used with the Large Field Objective and $J$, $H$ and
$K_{\mathrm{s}}$ filters for NIR imaging.

We took a set of very deep images in each filter in order to search for possible
counterpart candidates not detected by previous surveys, and to perform
accurate photometry of all candidates. Due to the nature of the high energy
source, photometric variability is expected, hence two sets of short,
contiguous exposures spanning 1--2 hours were also taken to construct optical
and NIR light curves respectively. We took also long exposure spectra of
USNO-B1.0~0636-0620933 in order to cover the whole optical spectral range with
a high S/N ratio. Tables~\ref{img} and \ref{spec} give the basic parameters
used for the observations. The standard reduction procedures for optical and
NIR images (sky/bias substraction
and flat-fielding) were used to obtain the final science images. The reduction
was performed using the IRAF package {\sc ccdred} (Tody \cite{Tod93}). The
spectra were pre-processed with MIDAS and then reduced using the Starlink
packages {\sc ccdpack} and {\sc figaro}.

\begin{table}
\begin{center}
\caption{Optical and NIR imaging log.}
\label{img}
\begin{tabular}{cccccc}
\hline
\hline
Filter & Exposure & \# frames & Filter & Exposure & \# frames \\
 & time (s) & & & time (s) & \\
\hline
$B$ & 300 &  1 & $J$              &  60 &   9\\
$V$ & 200 &  1 & $H$              &  60 &   9\\
$V$ &  10 & 90 & $K_{\mathrm{s}}$ &  60 &   9\\
$R$ & 200 &  1 & $K_{\mathrm{s}}$ &   2 & 366\\
\hline
\end{tabular}
\end{center}
\end{table}

\begin{table}
\begin{center}
\caption{Spectroscopy log.}
\label{spec}
\begin{tabular}{lccc}
\hline
\hline
Grating & Exposure & Wavelength & Resolution\\
or grism & time (s) & range (\AA)& (\AA)\\
\hline
\#3  & 150 & 3600--8400 & 7.9\\
\#5  & 400 & 3800--6700 & 4.5\\
\#6  & 300 & 6180--6860 & 1.2\\
\#12 & 700 & 4000--4920 & 2.6\\
\hline
\end{tabular}
\end{center}
\end{table}

\section{Data analysis and results}
\label{res}

\subsection{Astrometry}
\label{ast}

\begin{figure}
\resizebox{\hsize}{!}{\includegraphics{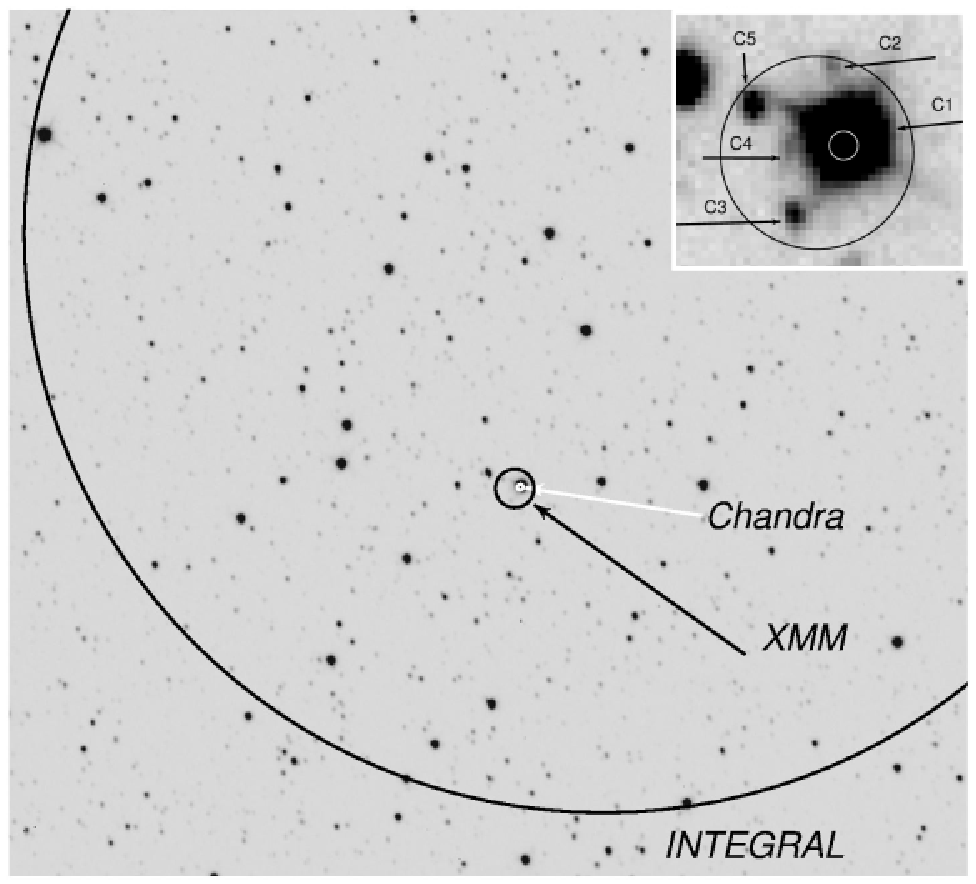}}
\caption{$K_{\mathrm{s}}$ band image of the field of
\object{IGR~J17544$-$2619}. North is up and East is to the left. The large
circular
section is the {\em INTEGRAL} error circle ($2\arcmin$ radius) for the source,
while the small circles are the {\em XMM-Newton} and {\em Chandra} error boxes
(black, $4\arcsec$ and white, $0.6\arcsec$ radius respectively). {\bf Inset:}
The five counterpart candidates (C1--C5) found inside the {\em XMM-Newton}
error box (black circle). The brightest candidate (C1), the only one allowed by
the {\em Chandra} error box (white circle), is
USNO-B1.0~0636-0620933/2MASS~J17542527$-$2619526, the candidate proposed by
Rodriguez (\cite{Rod03}). The image was processed to unveil the faintest
candidates, hence the intensity scale is nonlinear.}
\label{cand}
\end{figure}

We performed the astrometry of our images to determine the position in our
frames of the {\em XMM-Newton} and {\em Chandra} error boxes for
\object{IGR J17544$-$2619} and to search for optical/NIR counterparts within
them. We chose the best frame in each band for this purpose, and selected from
it a set of bright pointlike objects in uncrowded regions. We took the
coordinates of these objects from the USNO B1.0 and the 2MASS catalogs (Monet
et~al. \cite{Mon03}; Cutri et~al. \cite{Cut03}), and computed the
plate solution for each frame using the IRAF {\sc ccmap} task, obtaining an rms
error $<0.15\arcsec$, which is small enough for our purposes. In
Fig.~\ref{cand} we show the field of the source in the $K_{\mathrm{s}}$ band,
together with the {\em INTEGRAL}, {\em XMM-Newton} and {\em Chandra} error
circles. Inside the second one we found 5 counterpart candidates (C1--C5, their
positions are given in Table~\ref{posc}). The brightest candidate (C1) is the
one proposed by Rodriguez (\cite{Rod03}), the other four (including one
apparently extended object, C4) are only visible in the NIR and are extremely
faint, close to our detection limit. The {\em Chandra} error box clearly
excludes all the candidates but C1, hence we conclude that this is the correct
counterpart of \object{IGR~J17544$-$2619}. The three faint pointlike objects
are probably foreground dwarf stars, while the extended one might be a
background galaxy.

\begin{table}
\begin{center}
\caption{Positions of optical/NIR counterpart candidates of the high energy
source \object{IGR~J17544$-$2619}, and their distances to the brightest one.}
\label{posc}
\begin{tabular}{lcccl}
\hline
\hline
Id. & $\alpha$ (J2000) & $\delta$ (J2000) & Distance & Notes \\
 & (h\,m\,s) & ($\degr\,\arcmin\,\arcsec$) & to C1 ($\arcsec$) & \\
\hline
C1 & 17:54:25.27 & $-$26:19:52.7 & 0.0 & Bright \\
C2 & 17:54:25.32 & $-$26:19:49.7 & 3.1 & Faint, pointlike \\
C3 & 17:54:25.44 & $-$26:19:55.8 & 3.9 & Faint, pointlike \\
C4 & 17:54:25.46 & $-$26:19:53.2 & 2.6 & Faint, extended? \\
C5 & 17:54:25.57 & $-$26:19:51.2 & 4.3 & Faint, pointlike \\
\hline
\end{tabular}
\end{center}
\end{table}

\subsection{Spectroscopy}
\label{spe}

\begin{figure}
\resizebox{\hsize}{!}{\includegraphics{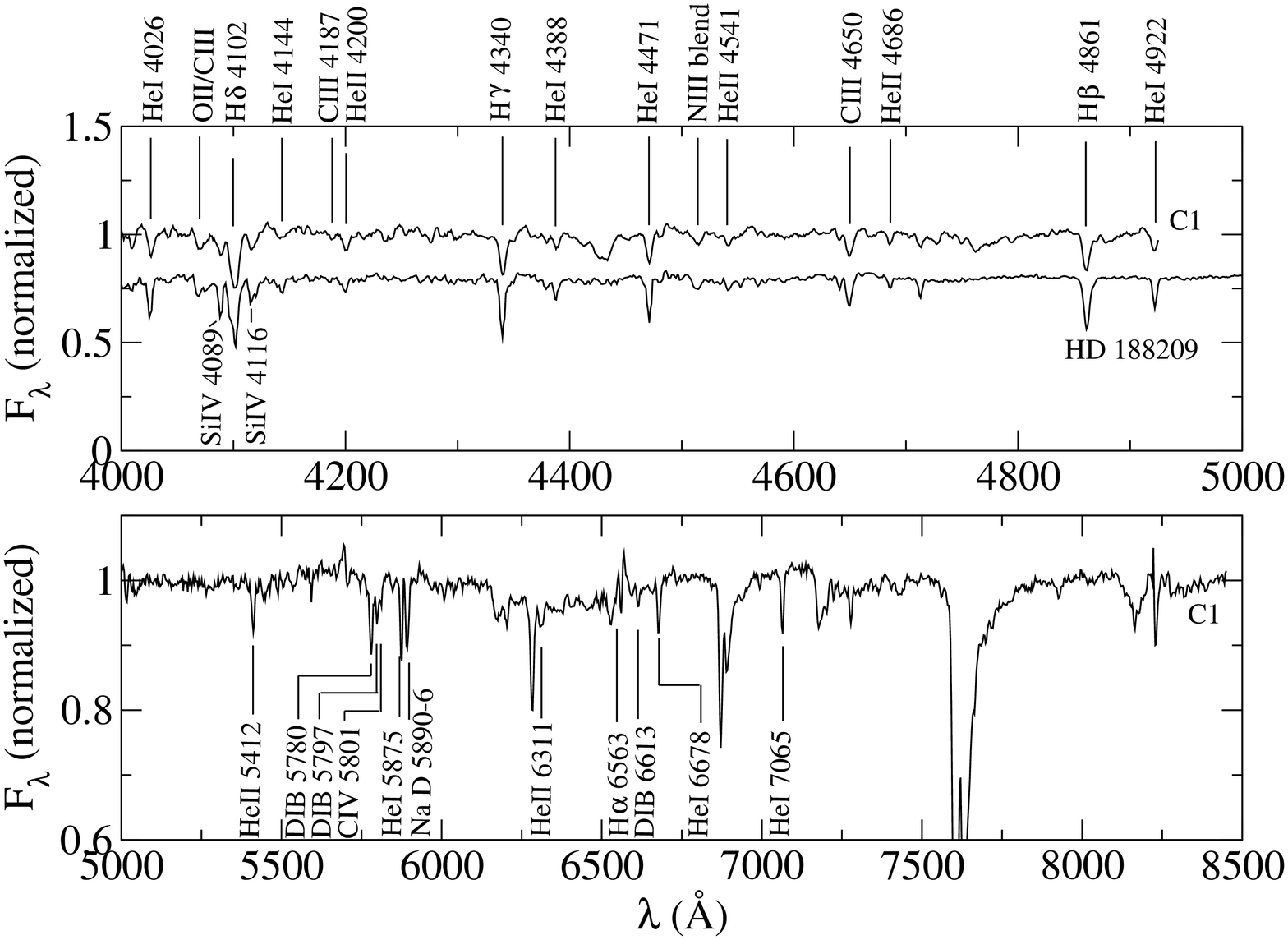}}
\caption{Optical spectrum of USNO-B1.0~0636-0620933 showing the identified
lines, among which we found strong \ion{He}{ii} lines typical of an O-type
star. {\bf Upper panel:} Blue spectrum of C1 taken with grism \#12 (upper
curve), and the standard O9Ib star \object{HD 188209} (lower curve). The high
degree of similarity between them supports our classification of C1 as an O9Ib
star. {\bf Lower panel:} Red spectrum of C1 between 5000~{\AA} and 8500~{\AA},
taken with grism \#3. The features at 6850~{\AA} and 7600~{\AA} are of telluric
origin.}
\label{spectra}
\end{figure}

Spectroscopic observations of candidate C1 enabled us to determine for the
first time its spectral type. In Fig.~\ref{spectra} we present the optical
spectra of this object taken with grisms \#12 and \#3, while in Table~\ref{lines} we list the spectral
features identified in it. The presence of intense \ion{He}{ii} lines in the
spectrum of C1 suggests an O spectral type. According to the classification
criteria of Walborn \& Fitzpatrick (\cite{Wal90}), the ratio
\ion{He}{ii}~4541~{\AA}/\ion{He}{i}~4471~{\AA} indicates that C1 is an O9 star.
The luminosity class of C1 could also be determined using the intensity of the
\ion{He}{ii}~4686~{\AA} as a luminosity indicator (Walborn \& Fitzpatrick
\cite{Wal90}). This line is very weak, but clearly visible in absorption in our
object, indicating a Ib luminosity class. Due to the low S/N ratio of the
spectrum, there is some uncertainty about this spectral classification, but the
O9Ib spectral type is fully supported by the lower resolution, higher S/N ratio
grism \#5 spectrum. Fig.~\ref{spectra} includes, for
comparison, the spectrum of the O9Ib star \object{HD 188209}, showing that the
general appearance of both spectra is similar. This classification confirms
that C1 is a blue supergiant and, via the calibration of Drilling \& Landolt
(\cite{Dri99}), provides the first determination of its mass,
25--28~$M_{\sun}$. Hence, we conclude that \object{IGR~J17544$-$2619} is a
HMXB. Another interesting feature of the C1 spectrum is the \ion{H}{$\alpha$}
line, that shows a P-Cygni type profile. This is the signature of mass loss
through a strong stellar wind, present in normal O-type stars.

\begin{table}
\begin{center}
\caption{Absorption lines identified in the spectrum of USNO-B1.0~0636-0620933,
and the atomic species that produce them.}
\label{lines}
\begin{tabular}{cclccl}
\hline
\hline
$\lambda_{\mathrm{obs}}$ & $\lambda_{\mathrm{lab}}$ & Species &
$\lambda_{\mathrm{obs}}$ & $\lambda_{\mathrm{lab}}$ & Species \\
({\AA}) & ({\AA}) & &
({\AA}) & ({\AA}) & \\
\hline
$4027 \pm 1$ & 4026.189 & \ion{He}{i} &
$4542 \pm 1$ & 4541.590 & \ion{He}{ii} \\
 & 4026.362 & \ion{He}{i} &
$4650 \pm 1$ & 4650.160 & \ion{C}{iii} \\
$4070 \pm 1$ & 4067.870 & \ion{C}{iii} &
 & 4651.350 & \ion{C}{iii} \\
 & 4068.970 & \ion{C}{iii} &
$4685 \pm 1$ & 4685.682 & \ion{He}{ii} \\
 & 4069.636 & \ion{O}{ii} &
$4861 \pm 1$ & 4861.332 & \ion{H}{$\beta$} \\
 & 4069.897 & \ion{O}{ii} &
$4921 \pm 1$ & 4921.929 & \ion{He}{i} \\
 & 4070.300 & \ion{C}{iii} &
$5412 \pm 3$ & 5411.524 & \ion{He}{ii} \\
$4089 \pm 1$ & 4088.863 & \ion{Si}{iv} &
$5881 \pm 3$ & 5875.618 & \ion{He}{i} \\
$4101 \pm 1$ & 4101.737 & \ion{H}{$\delta$} &
 & 5875.650 & \ion{He}{i} \\
$4115 \pm 1$ & 4116.104 & \ion{Si}{iv} &
 & 5875.989 & \ion{He}{i} \\
$4142 \pm 1$ & 4143.759 & \ion{He}{i} &
 & 5889.953 & \ion{Na}{i} \\
$4187 \pm 1$ & 4187.050 & \ion{C}{iii} &
 & 5895.923 & \ion{Na}{i} \\
$4200 \pm 1$ & 4199.830 & \ion{He}{ii} &
$6310 \pm 3$ & 6310.800 & \ion{He}{ii} \\
$4340 \pm 1$ & 4340.468 & \ion{H}{$\gamma$} &
$6564 \pm 1$ & 6562.817 & \ion{H}{$\alpha$} \\
$4388 \pm 1$ & 4387.928 & \ion{He}{i} &
$6679 \pm 3$ & 6678.149 & \ion{He}{i} \\
$4471 \pm 1$ & 4471.477 & \ion{He}{i} &
$7062 \pm 3$ & 7065.188 & \ion{He}{i} \\
 & 4471.688 & \ion{He}{i} &
 & 7065.719 & \ion{He}{i} \\
$4515 \pm 1$ & 4510.920 & \ion{N}{iii} &
 & & \\
 & 4514.890 & \ion{N}{iii} &
 & & \\
\hline
\end{tabular}
\end{center}
\end{table}

The spectrum of C1 shows also several diffuse interstellar bands (DIBs),
usually present in the spectra of sources with large extinction. The DIBs at
5780~{\AA}, 5797~{\AA} and 6613~{\AA} correlate with the
color excess $E(B-V)$ (e.g., Herbig \cite{Her93}; Cox et~al. \cite{Cox05}),
hence we can use them to determine the latter. In our spectra, the DIB at 
5797~{\AA} is blended with the \ion{C}{iv} 5801~{\AA} line, but the other two
DIBs are clearly resolved. The first one shows an equivalent width $W_{5780} =
1020 \pm 80$~m{\AA} implying, through the relationship given by Herbig
(\cite{Her93}), a color excess $E(B-V) = 1.97 \pm 0.15$. For the other one, our
measured equivalent width ($W_{6613} = 300 \pm 50$~m\AA) is contained in the
region where the relationship between $W$ and $E(B-V)$ is nonlinear. Figure 4
of Cox et~al. (\cite{Cox05}) shows that this value is consistent with $E(B-V) >
1$, the flattening of the relationship preventing any other meaningful
conclusion.

\subsection{Photometry}
\label{phot}

Differential aperture photometry of C1 was made using the IRAF package
{\sc apphot}. For each frame, we measured its instrumental optical and NIR
magnitudes $m_B$, $m_V$, $m_R$, $m_J$, $m_H$ and $m_{K_{\mathrm{s}}}$, together
with those of 18 comparison stars with known USNO~B1.0 and 2MASS magnitudes.
The Johnson $B$, $V$ and $R$ magnitudes of the comparison stars were computed
from their photographic IIIa-J (blue) and IIIa-F (red) magnitudes taken from
the USNO~B1.0 catalog and the transformations of Windhorst et~al.
(\cite{Win91}), while their $J$, $H$ and $K_{\mathrm{s}}$ standard magnitudes
were taken directly from the 2MASS catalog.

Linear fits of the standard optical magnitudes against corresponding
instrumental magnitudes present slopes consistent with unity within errors of a
few percent, showing that our instrumental magnitudes differ from the standard
system only in the zero points for the different bands. These were computed by
taking the mean of the differences $B-m_B$, $V-m_V$ and $R-m_R$ for the
comparison stars. Their uncertainties were determined by assuming that these
differences follow a normal distribution, and computing the error of the mean.
The
standard deviation of these differences is $\sim$0.2~mag, consistent with the
facts that USNO~B1.0 uncertainties are around 0.3~mag in the worst cases (Monet
et~al. \cite{Mon03}) and that the transformations between photographic
and Johnson systems add an error of $\sim$0.1~mag in quadrature (Windhorst
et~al. \cite{Win91}). Hence, our zero points have uncertainties of
$\sim$0.05~mag.

For NIR images, C1 was slightly in the nonlinear range of SOFI
CCDs, hence quadratic fits of standard 2MASS magnitudes $J$, $H$ and
$K_{\mathrm{s}}$ against corresponding instrumental magnitudes $m_J$, $m_H$ and
$m_{K_{\mathrm{s}}}$ respectively were used to derive the NIR magnitudes of C1.
The errors of the estimated magnitudes were computed from the full covariance
matrix of the fitted parameters, and take into account the uncertainties of
the 2MASS magnitudes of the comparison stars, our instrumental magnitude errors
being much smaller.

The optical and NIR magnitudes of C1 are shown in Table~\ref{magc}. The value
of $R$ is an upper limit as the image of C1 was slightly saturated in this
band. We performed the photometry for different aperture diameters $d$ from
1$\arcsec$ to 5$\arcsec$ in order to detect any flux contribution from the fainter
candidates. In all cases, the constancy of the C1 instrumental magnitudes for
$d > 2.5\arcsec$ up to an accuracy of 0.01~mag allows us to put conservative
lower limits for the combined magnitude of all of them (see Table~\ref{magc}).
The contribution of C2--C5 is thus negligible and we can ensure that it does
not contaminate our results on C1 at the 0.01~mag accuracy level.

\begin{table}
\begin{center}
\caption{Optical and NIR magnitudes of the counterpart candidates of the high
energy source \object{IGR~J17544$-$2619}. The lower limits stand for the
combined magnitudes of the four faint candidates C2--C5.}
\label{magc}
\begin{tabular}{lcccccc}
\hline
\hline
Id. & $B$ & $V$ & $R$ & $J$ & $H$ & $K_\mathrm{s}$ \\
 & $\pm 0.05$ & $\pm 0.05$ & & $\pm 0.02$ & $\pm 0.02$ & $\pm 0.02$ \\
 & (mag) & (mag) & (mag) & (mag) & (mag) & (mag) \\
\hline
C1 & 14.44 & 12.65 & $<$ 11.9 & 8.71 & 8.03 & 7.99 \\
C2--5 & $>$ 19.5 & $>$ 17.7 & $>$ 16.2 & $>$ 13.8 & $>$ 13.1 & $>$ 13.1 \\
\hline
\end{tabular}
\end{center}
\end{table}

Our broad-band magnitudes are in accordance with those listed in
Rodriguez (\cite{Rod03}), and also with the $B$ magnitude measured by {\em
XMM-Newton} (Gonz\'alez Riestra et~al. \cite{GRi04}). We note that our $J$ and
$H$ magnitudes differ (the first one marginally) from those given by 2MASS, our
estimates being lower than catalog values; but our $K_{\mathrm{s}}$ value
agrees well with 2MASS data ($J_{\mathrm{2MASS}} = 8.791 \pm 0.021$,
$H_{\mathrm{2MASS}} = 8.310 \pm 0.031$, $K_{\mathrm{s,2MASS}} = 8.018 \pm
0.026$). Because of the high accuracy of both 2MASS and our measurements, and
the fact that our determinations are based on the 2MASS magnitudes of the
comparison stars, we conclude that the differences might arise in a true
variation in the NIR flux of C1, brighter in September 2003 than in the epoch
of 2MASS observations (July 1998). This variation makes a stronger case for the
identification of C1 as the correct counterpart of \object{IGR~J17544$-$2619}.
We suggest that it might be related to the X-ray activity of the source. A more
detailed study through simultaneous multiwavelength campaigns both during
quiescence and activity would be important to confirm these points.

The combination of the spectral type of C1 with its photometry gave us a deeper
insight into the nature of the system. An O9Ib star has an intrinsic colour
$(B-V)_0 = -0.28 \pm 0.01$ (Schmidt-Kaler \cite{Sch82}), whereas we measure
$B-V = 1.79 \pm 0.10$, implying a reddening $E(B-V) = 2.07 \pm 0.11$. This
value is consistent with that found using the DIBs observed in the spectra
of C1. Hereafter we use the average of both, $E(B-V) = 2.02 
\pm 0.13$. Assuming
a standard extinction law ($R_V = 3.1$) gives a total visual extinction $A_V =
 6.26 \pm 0.40$. There is no reliable absolute visual magnitude calibration for
O9Ib stars alone, but as Ib stars are the least luminous objects of class I, we
take the mean value for {\em all} O9I stars, $M_V = -6.29$ (Martins et~al.
\cite{Mar05}), as an upper limit for the luminosity of C1. On
the other hand, the faintest O9I stars have $M_V \sim -5.6$ (Martins et~al.
\cite{Mar05}). With this interval for the absolute visual magnitude of
C1, we derive a range of 2.1--4.2~kpc for the distance of the system, far
smaller than the 8.5~kpc computed by Smith (\cite{Smi04}) using USNO B1.0
photometry. The disagreement arises from the lack of accuracy of USNO
magnitudes. This result is very robust, even if C1 were among the brightest O9I
stars, the distance would be still smaller than $\sim5$~kpc, and if a value
$R_V > 3.1$ were used, the distance would decrease. Hence, the most probable
localization of \object{IGR~J17544$-$2619} is between 2--4~kpc of the Sun, in
the Scutum--Crux arm of the Milky Way (see Vall\'ee \cite{Val05} for a model of
the Milky Way spiral arms). We note that small distances were also obtained for
\object{XTE~J1739$-$302} (1.8--2.9~kpc; Negueruela et~al. \cite{Neg06}) and
\object{IGR~J16318$-$4848} ($<$~4~kpc; Filliatre \& Chaty \cite{Fil04}).

The observed extinction implies a hydrogen column density $N_{\mathrm{H}} =
1.2 \times 10^{22}$~cm$^{-2}$, close to the interstellar value (1.4$\times
10^{22}$~cm$^{-2}$, Dickey \& Lockman \cite{Dic90}). Our value is consistent
with those obtained from {\em Chandra} observations ($1.36 \pm 0.22 \times
10^{22}$~cm$^{-2}$) and lower than {\em XMM-Newton} results (1.9--4.3$\times
10^{22}$~cm$^{-2}$), showing that the higher $N_{\mathrm{H}}$ values are
obtained during X-ray activity, $N_{\mathrm{H}}$ decreasing during quiescence.
This suggests the existence of a highly variable circumstellar medium around
the system, whose presence is related to the high energy activity of the
source. We note that the possibility of the extinction being always
completely interstellar is difficult to reconcile with the variable
$N_{\mathrm{H}}$ suggested by the combination of X-ray and our observations.

\begin{figure}
\resizebox{\hsize}{!}{\includegraphics{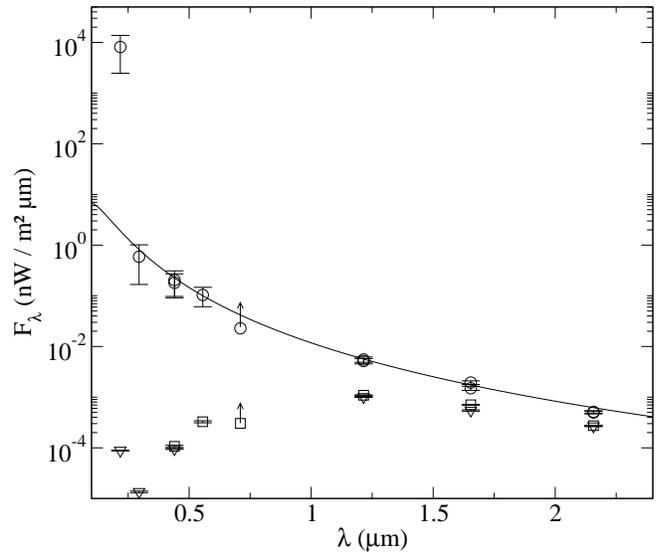}}
\caption{Spectral energy distribution $F_\lambda (\lambda)$ of the source.
Squares represent the fluxes observed by us and triangles those from
{\em XMM-Newton} and 2MASS data. Circles correspond to
extinction corrected fluxes. The solid line shows the SED of a blackbody at the
effective temperature of an O9Ib star ($T_{\mathrm{eff}} = 3.1 \times 10^4$~K),
normalized to have the same unabsorbed $V$ magnitude of the source.}
\label{sed}
\end{figure}

The photometry obtained allowed us to construct for the first time a spectral
energy distribution (SED) of the source, which we show in Fig.~\ref{sed}.
Squares in this figure represent our measurements of the observed
flux of the source $F_\lambda$ as a function of wavelength $\lambda$, while
triangles represent {\em XMM-Newton} and 2MASS data. Using the estimated 
visual extinction and its relationship to the extinction in other bands (Mathis
\cite{Mat99}), we computed the latter and corrected the SED for its effects
(circles in Fig.~\ref{sed}). We also constructed the SED of an O9Ib
star normalized to have the same unabsorbed $V$ magnitude of C1, using its
intrinsic color indices (Drilling \& Landolt \cite{Dri99}; Tokunaga
\cite{Tok99}). We found that it is very well fitted by a blackbody with a
temperature equal to the effective temperature of an O9I star 
($T_{\mathrm{eff}} = 3.1 \times 10^4$~K, Martins et~al. \cite{Mar05}), 
normalized in the same way (solid line in Fig.~\ref{sed}). We note the very 
good agreement between the two unabsorbed SEDs in the optical, NIR and part of
the UV, confirming our assumption of standard reddening . The lack of detection
of C1 in the $UVM2$ band ($\lambda = 0.234$~$\mu$m, Gonz\'alez-Riestra et~al.
\cite{GRi04}) is consistent with the blackbody spectrum and the amount of
extinction derived by us. However, the detection at 0.218~$\mu$m ($UVW2$ band)
is puzzling, because the absorbed flux should be more than two magnitudes below
the limiting magnitude of the {\em XMM-Newton} OM for this wavelength (there is
a strong enhancement in the extinction curve at this wavelength, the 2175~{\AA}
bump). A peculiar value for the extinction at 0.218~$\mu$m (e.g., due to the
depletion of the material that produces the 2175~{\AA} bump) could be one
solution, but it is improbable given the good agreement in other wavelengths
and the fact that a reduction of $\sim$9~mag in the extinction in the $UVW2$
band is needed to restore the agreement with the blackbody spectrum. On the
other hand, if the computed absorption is correct, the measured magnitude
($m_{UVW2} = 14.49 \pm 0.02$) implies that the unabsorbed flux at this
wavelength is more than three orders of magnitude greater than the blackbody
estimate. This disagreement could be explained if the source were variable in
the UV, because the observations in the $UVW2$ and $UVW1$ ($\lambda =
0.295$~$\mu$m) bands were separated by 6 days. If this is the case, it would be
interesting also to determine if the variability is connected to the X-ray
behaviour. Simultaneous multiwavelength campaigns are clearly needed to
investigate this issue and determine the astrophysical processes which operate
in this source.

\subsection{Light curves}
\label{lc}

\begin{figure}
\resizebox{\hsize}{!}{\includegraphics{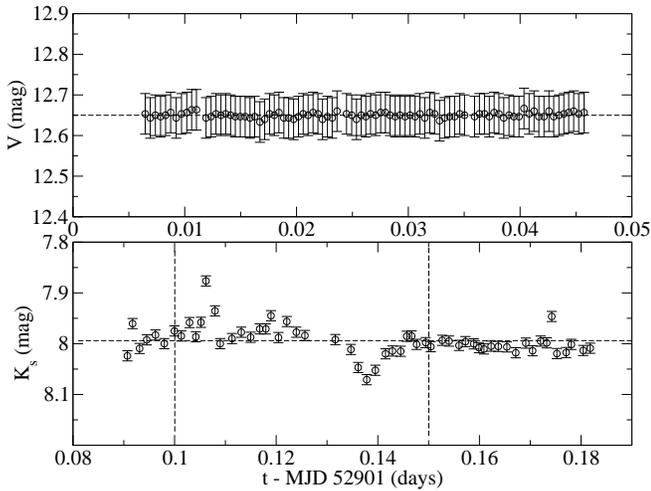}}
\caption{Optical and NIR light curves of C1. Abscissas are given as Modified
Julian Dates (MJD), MJD 52901 being September 19, 2003, 0$^{\mathrm{h}}$ UT.
{\bf Upper panel:} The $V$ band light curve is extremely flat, showing a
constant contribution from the O star. {\bf Lower panel:} The $K_{\mathrm{s}}$
band light curve shows erratic variations around $t -
\mathrm{MJD~52901}$~$\sim$~0.10--0.15~days (between the two dashed lines), note
the two brightenings ($\sim$0.10~mag at 0.105~days and $\sim$0.05~mag at
0.12~days) and the fading ($\sim$0.10~mag at 0.14~days). The differences with
the $V$ band curve suggest that they are probing different parts of the system,
and supports the picture of the existence of circumstellar material around the
star.}
\label{curves}
\end{figure}

Using the 90 $V$-band and the 366 $K_{\mathrm{s}}$-band short exposures (see
Table~\ref{img}), we constructed optical/NIR light curves of C1 spanning 3.4~ks
and 8~ks respectively (Fig.~\ref{curves}). The NIR data were averaged in groups
of 4,
which results in a lower uncertainty of $\sim$0.01~mag. The $V$ light curve is
flat and completely featureless down to our accuracy level of 0.05~mag. The
$K_{\mathrm{s}}$ light curve, instead, shows an erratic behaviour barely above
our accuracy limits of 0.01~mag. A few sporadic events are seen around $t -
\mathrm{MJD~52901}$~$\sim$~0.10--0.15~days, including two short brightenings of
approximately 0.05--0.10~mag  and an abrupt fading of about 0.10~mag,
reminiscent of an eclipse. No periodicity is apparent in our light curves.

Although the optical and NIR light curves are not simultaneous, they are close
in time, hence their different behaviours suggest that they are probing
different parts of the system. The extreme flatness of the optical light curve
contrasts with the multiple short erratic variations found in the NIR light
curve. The $V$ light curve is probably reflecting the behaviour of the
O9Ib star, as it is the brightest source in this spectral domain. We note 
that the constant light curve observed is not in contradiction with the
fact that most O stars are microvariable in $V$ (e.g., Balona \cite{Bal92}),
because their amplitudes are $<0.1$~mag, lower than or of the same magnitude of
our accuracy ($\pm0.05$~mag). The NIR variations, that occur on timescales of
the order of hours, could be generated by a different component. A possible
candidate is a variable amount of warm material (either circumstellar or in an
accretion disc) emitting in the NIR. We note that \object{IGR~J17544$-$2619}
was recently detected in the
medium IR (8.59~$\mu$m and 11.26~$\mu$m) by the VISIR instrument on VLT (Chaty
et~al., in prep.). We also note that the erratic variability observed in 
$K_{\rm s}$ could explain the differences between our NIR magnitudes and 2MASS
values. Longer light curves in several optical and IR bands together with 
optical/NIR spectra would be needed to address this issue.

\section{Discussion}
\label{disc}

Using optical and NIR imaging of the field of \object{IGR~J17544$-$2619} we
identified the optical/NIR counterpart of this high energy source to be
\object{USNO-B1.0~0636-0620933}/\object{2MASS~J17542527$-$2619526}.
We have obtained strong evidence for \object{IGR~J17544$-$2619} being a high
mass X-ray binary. Although our data do not give any clue about the accretor
in the HMXB, they can be obtained from the X-ray data. The peak flux density
scaled to 1~kpc (0.5--1.9~Crab, for our distance range of 2.1--4.2~kpc)
and the spectral index obtained by {\em Chandra} are consistent with those of
black holes in the hard state. Applying the relationship of Gallo et~al.
(\cite{Gal03}), we obtain an expected radio flux of $\sim$10--35~mJy for
this source if the accretor is a black hole. Hence, the lack of detection of
\object{IGR~J17544$-$2619} in radio (3$\sigma$ upper limit of 7~mJy at
0.61~GHz, Pandey et~al.
\cite{Pan06}) suggests that the compact object is rather a neutron star. This
agrees with the results of in't Zand (\cite{Zan05}), who shows that the X-ray
spectrum of \object{IGR~J17544$-$2619} in quiescence can be satisfactorily
fitted by a neutron star model with a luminosity of $\sim 5 \times
10^{32}$~erg~s$^{-1}$.

For the mass donor, our results show that it is an O9Ib blue supergiant with a
mass of 25--28~$M_{\sun}$. Hence, at its birth the object should have been an
O+O binary, a rare system due to the high masses of its components and their
short lifetimes (a 25~$M_{\sun}$ star lives only $\sim$7~Myr in the main
sequence). This characteristic is consistent with the fact that we find
\object{IGR~J17544$-$2619} in a star-forming region of the Galaxy (the
Scutum-Crux arm at $\sim$3--4~kpc). It is not surprising then to find other
systems with similar properties in nearby regions (Scutum-Crux and Norma arms).
Although massive binaries undergo mass-transfer phases before the primary
evolves into a compact object, recent results suggest that the maximum mass
gained by the secondary could be as small as 10\% of the mass transferred
(e.g., Petrovic et~al. \cite{Pet05}). This suggests that the mass of the
primary should have been $\gtrsim 25$~$M_{\sun}$, and hence that such massive
stars are not necessarily constrained to evolve into black holes.

Our data, combined with X-ray observations, suggest that this system is
embedded in circumstellar material. We derive a moderately high hydrogen column
density $N_{\mathrm{H}} = 1.2 \times 10^{22}$~cm$^{-2}$, that agrees well with
{\em Chandra} results and is lower than that derived from {\em XMM-Newton}
data (Gonz\'alez-Riestra et~al. \cite{GRi04}; in't Zand \cite{Zan05}). This
confirms that the extinction of the system is variable, and hence it can not be
completely interstellar. The existence of NIR variability with no optical
correlation shown by our light curves could be explained if we assume that
optical and NIR emission arise in different regions of the system. The optical
radiation comes from the blue supergiant, hence at least part of the
NIR should come from another region, the circumstellar environment being a good
candidate. This scenario is interesting regarding the recent detection of
\object{IGR~J17544$-$2619} in the medium IR.

Our analysis, together with other recents studies (Smith \cite{Smi04}; in't
Zand \cite{Zan05}; Negueruela et~al. \cite{Neg06}; Smith et~al. \cite{Smi06})
suggest that the
group of the sources discovered by {\em INTEGRAL} and {\em RXTE} observatories
close to the direction of the galactic center is not homogeneous, it rather
shows two distinct subgroups. The first one comprises HMXBs with persistent
X-ray emission, large circumstellar absorption and supergiant mass donors
(e.g., \object{IGR~J16318$-$4848}, Filliatre \& Chaty \cite{Fil04}). The second
group, dubbed {\em supergiant fast X-ray transients} (SFXT) by Negueruela
et~al. (\cite{Neg06}) and Smith et~al. (\cite{Smi06}), includes also
supergiant HMXBs, but with fast
($\sim$hours) transient X-ray emission, very low quiescence X-ray luminosity
($\leq 10^{34}$~erg~s$^{-1}$), and moderate to high variable X-ray extinction
originated in the circumstellar environment. \object{IGR~J17544$-$2619} belongs
to this second group, together with \object{XTE~J1739$-$302} (the prototype of
the group, Negueruela et~al. \cite{Neg06}), \object{IGR~J16465$-$4507},
\object{AX~1845.0$-$0433} and possibly \object{XTE~J1901+014}. SFXTs include
also at least three other sources, \object{SAX~J1818.6$-$1703},
\object{AX~J1749.1$-$2733} (in't Zand {\cite{Zan05}) and
\object{IGR~J11215-5952} (Sidoli et~al. \cite{Sid06}).

As pointed out by in't Zand ({\cite{Zan05}) and Negueruela et~al.
(\cite{Neg06}), the physical mechanism driving the fast outbursts in SFXTs
would be related neither to the compact object, as the group includes both
neutron star and black hole HMXBs, nor to the low luminosity, as fast outbursts
were observed also in \object{Vela~X-1} and \object{Cygnus X-1}. The accretion
mechanism is then the most probable cause of the short duration of the
outbursts. As supergiant HMXBs are believed to be powered by accretion from the
strong wind of the secondary, the variation of the mass flux of the wind is
probably the cause of the outbursts. The changing distance of the mass donor in
an excentric orbit is an attractive posibility for explaining this mass flux
variation, as it could also explain the low quiescent luminosity and small duty
cycle of the X-ray source if the O star spends most of the orbital period far
away from the accretor. We point out that this is consistent with the fact
that recurrent outbursts have been recently found in an SFXT
(\object{IGR~J11215-5952}), with a period of 329~days (Sidoli et~al. \cite{Sid06}). On the other hand, local inhomogeinities (clumps) in
the wind were proposed as an alternative to explain the fast outbursts, as they
are individually accreted by the compact object (in't Zand \cite{Zan05}). We
note that, if as we propose, the variability of the circumstellar medium and
the X-ray activity are correlated, then individual clump accretion could not
explain the whole phenomenology. A global wind density variation (whether it
is composed of clumps or not) would be a better explanation for the outbursts,
provided that it lasts for a time of the order of a few hours.

Finally, we point out that a more detailed monitoring of this system would be
important to completely unveil its properties and hence improve our knowledge
on this kind of high energy sources. Particularly, simultaneous multiwavelength
campaigns would be needed to precisely measure the SED both in quiescence and
during activity, establish the existence of circumstellar material beyond any
doubt, and determine its properties. The analysis of both long and short-term
variability (including the search for a binary period and wind velocity
variability) and their connection to high energy activity would also shed light
over the physical mechanisms involved in the behaviour of these systems.

\begin{acknowledgements}

We would like to thank Dr. Marc Rib\'o for useful discussions and a careful
reading of the manuscript, and the anonymous referee for suggestions which
greatly improved this paper.
IN is a researcher of the program {\em Ram\'on y Cajal}, funded by the Spanish
Ministerio de Ciencia y Tecnolog\'{\i}a and the University of Alicante, with
partial support from the Generalitat Valenciana and the European Regional
Development Fund (ERDF/FEDER). This research is partially supported by the
Spanish MCyT under grant AYA2002-00814.
This publication makes use of data products from the
Two Micron All Sky Survey, which is a joint project of the University of
Massachusetts and the Infrared Processing and Analysis Center / California
Institute of Technology, funded by the National Aeronautics and Space
Administration and the National Science Foundation. 
This research has made use of the SIMBAD database and VizieR Service
operated at CDS, Strasbourg, France, and of NASA's Astrophysics Data System
Bibliographic Services.

\end{acknowledgements}


\begin{thebibliography}{}

\bibitem[1992]{Bal92}
Balona, L.~A., 1992,
\mnras, 254, 403

\bibitem[2005]{Cox05}
Cox, N.~L.~J., Kaper, L., Foing, B.~H., \& Ehrenfreund, P., 2005,
\aap, 438, 187

\bibitem[2003]{Cut03}
Cutri, R.~M., Skrutskie, M.~F., van Dyk, S., et al. 2003,
2MASS All-Sky Catalog of Point Sources, University of Massachusetts and
Infrared Processing and Analysis Center, (IPAC / California Institute of
Technology). Vizier online catalog II/246.

\bibitem[1990]{Dic90}
Dickey, J.~M., \& Lockman, F.~J. 1990,
\araa, 28, 215

\bibitem[1999]{Dri99}
Drilling, J.~S., \& Landolt, A.~U. 1999,
in ``Allen's Astrophysical Quan\-ti\-ties'', ed. A.N. Cox, Springer, 381

\bibitem[2004]{Fil04}
Filliatre, P., \& Chaty, S. 2004,
\apj, 616, 469

\bibitem[2003]{Gal03}
Gallo, E., Fender, R.~P., \& Pooley, G.~G. 2003,
\mnras, 344, 60

\bibitem[2003]{Gre03}
Grebenev, S.~A., Lutovinov, A.~A., \& Sunyaev, R.~A. 2003
ATel, 192

\bibitem[2004]{Gre04}
Grebenev, S.~A., Rodriguez, J., Westergaard, N.~J., Sunyaev, R.~A., \&
Oosterbrock, T. 2004
ATel, 252

\bibitem[2004]{GRi04}
Gonz\'alez-Riestra, R., Oosterbroek, T., Kuulkers, E., Orr, A., \& Parmar,
A.~N. 2004,
\aap, 420, 589

\bibitem[1993]{Her93}
Herbig, G.~H., 1993,
\apj, 407, 142

\bibitem[2005]{Zan05}
in't Zand, J.~J.~M. 2005,
\aap, 441, L1

\bibitem[2005]{Kuu05}
Kuulkers, E. 2005,
AIP Conference Proceedings, 797, 402

\bibitem[2005]{Mar05}
Martins, F., Schaerer, D., \& Hillier, D.~J. 2005,
\aap, 436, 1049

\bibitem[2005]{Mas05}
Masetti, N., Pretorius, M.~L., Palazzi, E., et~al. 2005,
\aap, in press, arXiv:astro-ph/0512399

\bibitem[1999]{Mat99}
Mathis, J.~S. 1999,
in ``Allen's Astrophysical Quan\-ti\-ties'', ed. A.N. Cox, Springer, 523

\bibitem[2003]{Mon03}
Monet, D.~G., Levine, S.~E., Canzian, B., et~al. 2003,
\aj, 125, 984

\bibitem[2004]{Neg04}
Negueruela, I. 2004,
in ``The Many Scales of the Universe - JENAM 2004 Astrophysics Reviews'', eds.
J.~C. del Toro Iniesta, et~al., Proc of the Joint European and Spanish
Astronomical Meeting, Granada, Spain, September 2004, arXiv:astro-ph/0411759

\bibitem[2005]{Neg05}
Negueruela, I., Smith, D.~M., \& Chaty, S. 2005
ATel, 429

\bibitem[2006]{Neg06}
Negueruela, I., Smith, D.~M., Harrison, Th.~E., \& Torrej\'on, J.~M. 2006
\apj, 638, 982

\bibitem[2006]{Pan06}
Pandey, M., Manchanda, R., Rao, A.P., Durouchoux, P. \& Ishwara-Chandra, C.~H.
2006,
\aap, 446, 471

\bibitem[2005]{Pet05}
Petrovic, J., Langer, N., \& van der Hutch, K.~A. 2005,
\aap, 435, 1013

\bibitem[2003]{Rod03}
Rodriguez, J. 2003
ATel, 194

\bibitem[1982]{Sch82}
Schmidt-Kaler, Th. 1982,
in ''Landolt-B\"ornstein New Series'', Group VI, vol. 2b, ed. K. Schaifers,
\& H.H. Voigt (Springer-Verlag), 1

\bibitem[2006]{Sid06}
Sidoli, L., Paizis, A., \& Mereghetti, S. 2006,
\aap, in press, arXiv:astro-ph/0603081

\bibitem[2004]{Smi04}
Smith, D.~M. 2004,
ATel, 338

\bibitem[2006]{Smi06}
Smith, D.~M. et~al. 2006,
\apj, 638, 974

\bibitem[2003]{Sun03}
Sunyaev, R.~A., Grebenev, S.~A., Lutovinov, A.~A., et~al. 2003,
ATel, 190

\bibitem[1993]{Tod93}
Tody, D. 1993,
in ``Astronomical Data Analysis Software and Systems II'', A.S.P. Conference
Ser., Vol 52, eds. R.~J. Hanisch, R.~J.~V. Brissenden, \& J. Barnes, 173

\bibitem[1999]{Tok99}
Tokunaga, A.~T. 1999,
in ``Allen's Astrophysical Quan\-ti\-ties'', ed. A.N. Cox, Springer, 143

\bibitem[2005]{Val05}
Vall\'ee, J.~P. 2005,
\aj, 130, 569

\bibitem[1990]{Wal90}
Walborn, N.~R., \& Fitzpatrick, E.~L.. 1990,
\pasp, 102, 379

\bibitem[2004]{Wij03}
Wijnands, R. 2003
ATel, 191

\bibitem[1991]{Win91}
Windhorst, R.~A., Burstein, D., Mathis, D.~F., et.~al. 1991
\apj, 380, 362

\end{thebibliography}
\end{document}